\documentclass[12pt]{iopart}
\newcommand{\bk}{{\bf k}}
\newcommand{\mbr}{{\mathbf{r}}}

\newcommand{\rmc}{{\mbox{c}}}
\newcommand{\rmsc}{{\mbox{sc}}}

\newcommand{\beq}{\begin{equation}}
\newcommand{\eeq}{\end{equation}}
\newcommand{\bay}{\begin{eqnarray}}
\newcommand{\eay}{\end{eqnarray}}

\begin{document}
\title[The sharp screening of the Coulomb problem]{
The impact of sharp screening on the Coulomb scattering problem in three dimensions}
\author{S L Yakovlev$^1$, M V Volkov${^{2,3}}$, E Yarevsky$^1$,  N Elander$^3$
 }

\address{$^1$ Department of Computational
Physics, St Petersburg State University, 198504 St Petersburg, Russia }
\address{$^2$ Department of Quantum Mechanics, St Petersburg State University, 198504 St Petersburg, Russia}
\address{$^3$ Department of Physics, Stockholm University, Alba Nova University Center, SE 106 91, Stockholm, Sweden}

\ead{yakovlev@cph10.phys.spbu.ru}
\begin{abstract}
The scattering problem for two particles interacting via the Coulomb potential is examined for the case
where the potential has a sharp cut-off at some distance.
The problem is solved for two complimentary situations, firstly when the interior part of the Coulomb potential is left
in the Hamiltonian and, secondly, when the long range tail is
considered as the potential.
The partial wave results are summed up to obtain the wave function in three dimensions. It is shown that in the
domains where the wave function is expected to be proportional to the
known solutions,
the proportionality is given 
by an operator acting on the angular part of the wave function. The explicit representation for this operator is obtained
in the basis of Legendre polynomials. We proposed a driven
Schr\"odinger equation including an inhomogeneous term of the finite range with purely outgoing asymptotics
for its solution in the case of the three dimensional
scattering problem with long range potentials.

\end{abstract}
\pacs{03.65.Nk}
\submitto{\JPA}
\maketitle

\section{Introduction}
\subsection{Background}
The Coulomb force is the dominating interaction in atomic and molecular physics. It is therefore the underlying force in
chemistry and biology. Describing reactions with charged particles is thus an
essential task for theoretical atomic and molecular physics as well as in chemistry.
Despite this fundamental importance, solving the Schr{\"o}dinger equation for even a three-body
problem is a very difficult task. The asymptotic form of the wave function for three-body breakup is
known to have very complicated structure.
A recent review \cite{R:Rshakeshaft09}
gave an overview of the most important articles which deal with the scattering problem of charged particles.
The majority of these  methods focus on solving the three-body problem
without explicit knowledge of the three-body breakup asymptotics.
Inspired by the methods of
\cite{NutCoh} and \cite{R:Rescigno97,R:McCurdy04}, in two recent studies we presented a
new and rigorous method which can be used to solve the Coulomb scattering problem by using exterior complex scaling
\cite{VEYY, ElanderFBS}. The former analyzes the single channel two-body problem
while the latter indicates how the same formalism can be generalized to the full multi-channel three-body problem.
In this contribution we demonstrate the construction of the formal
part of the three-dimensional formulation of the three-body problem outlined in \cite{ElanderFBS} by studying the
three-dimensional two-body problem.

\subsection{Definition of the problem}
 The scattering solution to the Schr\"odinger equation
\begin{equation}
\left[ -\Delta_{\mbr}  +  V_{\rmc}(r) -k^2\right]\psi_{\rmc}(\mbr,\bk)=0
\label{SE}
\end{equation}
for the Coulomb potential
$V_{\rmc}(r)=2\eta k/r$
has the well known form \cite{T, Gordon, Messiah}

\begin{equation}
\psi_{\rmc}(\mbr,\bk)=\Gamma(1+\rmi\eta)\rme^{-\pi\eta/2}\rme^{\rmi\mbr\cdot\bk}
{_1}F_{1}(-\rmi\eta,1,\rmi(rk-\mbr\cdot\bk)).
\label{CWF}
\end{equation}
Here the vectors $\mbr$ and $\bk$ describe the position  and  the incident momentum.
Their magnitudes are denoted by $r$ and $k$. $\Gamma$ and $_1F_1$  are the Gamma function and the Confluent
Hypergeometric function, respectively. Another way of representing the solution
$\psi_{\rmc}(\mbr,\bk$) 
is the partial wave decomposition \cite{Messiah}
\begin{equation}
\psi_{\rmc}(\mbr,\bk)= \frac{1}{kr}\sum_{\ell=0}^{\infty}(2\ell+1)\rmi^{\ell}\,\rme^{\rmi\sigma_{\ell}}
F_{\ell}(\eta,kr)P_{\ell}(\cos \theta).
\label{pwcwf}
\end{equation}
Here $\cos \theta=\mbr\cdot\bk/rk$, $\sigma_{\ell}=\arg \Gamma(\ell+1+\rmi\eta)$ represents the Coulomb phase shift,
and $P_{\ell}$ is the $\ell$th Legendre polynomial. The regular Coulomb
wave function $F_{\ell}(\eta,kr)$ \cite{Abram} obeys the partial wave equation 
\begin{equation}
\left( -\frac{\rmd^2}{\rmd r^2} +\frac{\ell(\ell+1)}{r^2}+\frac{2\eta k}{r}-k^2\right)F_{\ell}(\eta,kr)=0
\label{pwce}
\end{equation}
and the boundary condition $F_{\ell}(\eta,0)=0$ at the origin.
With this choice of coordinates, the wave function depends on the triad $r,u=\cos\theta,k$,  such that
$\psi_{\rmc}=\psi_{\rmc}(r,u,k)$.

Although the explicit representations (\ref{CWF},\ref{pwcwf})
have been known since the very earliest stages of Quantum Mechanics \cite{T, Gordon},  the various approximations, which are
based on the procedures necessary for suppression of the long range tail of the Coulomb potential, have been studied for many decades.
The total number of publications on this subject is
enormous, and so here we quote only those few which focus on the principal aspects of the problem
\cite{T, Gordon, Ford, Taylor, Gorshkov}. Such procedures, called screening,  are of
substantial interest in view of their application to the scattering problem for more than two particles
since in that case the exact solution of the Coulomb problem is not available \cite{AGS,
Fonseca}.

In principal, two kinds of screening procedures exist, i.e., the sharp and the soft.
The sharp, which cuts off  the Coulomb potential beyond some radius $R$, leads to the finite range
potential 
\begin{equation}
V_R(r)=V_{\rmc}(r)\Theta(R-r).
\label{V_R}
\end{equation}
Here $\Theta$ is the standard Heaviside-function such that $\Theta(t)=1(0),\ t\ge 0 (t<0)$.
Soft screening methods imply multiplication of $V_{\rmc}$ by a smooth screening factor as,
for instance,  is done in the Yukawa potential $W_{\rho}(r)=V_{\rmc}(r)\rme^{- r/\rho}$.
The Coulomb wave function can be obtained in the limit $\rho\to\infty$ from the solutions of the
Schr\"odinger equation with the potential $W_{\rho}$.
A regularizing factor is needed to obtain the correct limit. For the three dimensional Yukawa potential this factor is
known analytically \cite{Gorshkov}.

The situation with the potential $V_R$ is more delicate. The representation for the Coulomb wave function through
the solution for the screened potential $V_R$ can easily be obtained at finite values
of $R$ for partial wave components \cite{Gordon, Ford, Taylor}. The regularizing factor in this case is $\ell$-dependent.
The infinite sum over $\ell$ should be computed to obtain the
solution in three dimensions. An accurate analysis of the asymptotics of the partial wave series for
the scattering amplitude as $R\to\infty$ was performed in \cite{Taylor} on the
basis of distribution theory. That led the authors to the commonly accepted asymptotic regularizing factor
$\rme^{-\rmi2\eta\log2kR}$ for the scattering amplitude for the $V_R$ potential. No
extra term possessing singularities in the forward (backward) scattering direction were observed in \cite{Taylor},
since the test functions used for the partial series summations were
assumed to be vanishing in
those directions. Actually, this requirement is not necessary.
The problem of deriving the correct
three dimensional expression for the wave function, which explicitly includes the Coulomb wave
function $\psi_{\rmc}$, was not the focus of the papers \cite{Ford, Taylor}. The paper \cite{Glockle}
attempted to solve the problem in three dimensions directly by solving the three
dimensional Lippmann-Schwinger equation with the potential $V_R$. However, the derivations made in \cite{Glockle}
have been performed only for a particular value of the coordinate $r=0$. It was not
proven that the solution obtained in \cite{Glockle} is valid for all values of $r$. In the comment
\cite{Popov} it was shown by direct calculations that the three-dimensional result of
\cite{Glockle} is erroneous.  A further discussion of the results of \cite{Glockle} and their relation to the results of
\cite{Taylor} can be found in \cite{Deltuva}.

The, to date, open situation of the cut-off Coulomb problem coupled with our own interests
\cite{VEYY}, related to the application of the complex rotation method for calculating the
scattering states in the system of particles with long range interactions, stimulated this research.
The present paper is devoted to studying the scattering problem for a
sharp cut-off Coulomb potential $V_R$ and its complement 
\begin{equation}
V^R=V_{\rmc}-V_R.
\label{V^R}
\end{equation}
In section two the partial wave equations are solved for a $V_R$ potential by the conventional matching procedure at the point
$r=R$ and the infinite sum over angular momenta $\ell$
is then evaluated in order to obtain the solution to the three dimensional Schr\"odinger equation.
The main result of this paper is that we prove in the region $r<R$ the wave function for the potential
$V_R$ in three dimensions is given by the action of an operator on the Coulomb wave
function. This operator  acts over the angular coordinate.
The asymptotics of this operator is evaluated as $R\to \infty$.
Our derivation supports the form of the regularizing factor for the wave function which was proposed in \cite{Ford}
for three dimensions but without a detailed proof. An extra term is found in the asymptotics of the scattering
amplitude which possesses fast oscillations as a function of $R$ and
delta-functional singularity in the forward scattering direction.
In section three we present the solution of the scattering problem for the Schr\"odinger equation with the $V^R$ potential.
To the best of the authors' knowledge
this represents the first time that this has been reported.
This solution is used in section four to construct the three dimensional driven Schr\"odinger equation with
the finite range potential
$V_R$ in the inhomogeneous term. As we demonstrated in our recent paper \cite{VEYY},
this equation is ideally suitable for applying the complex rotation method to solve the scattering problem
with long range interactions.

\section{The scattering problem for the potential $V_R$}
The partial wave equation
\begin{equation}
\left( -\frac{\rmd^2}{\rmd r^2}+\frac{\ell(\ell+1)}{r^2}+V_R(r)-k^2 \right)v_{\ell}(r,k)=0
\label{pweV_R}
\end{equation}
with the boundary condition $v_{\ell}(0,k)=0$ and the asymptotics as $r\to\infty$
\begin{equation}
v_{\ell}(r,k)\sim {{\hat {j_\ell}}}(kr) + A_{R  \ell}\, {\hat h}^+_{\ell}(kr)
\label{vas}
\end{equation}
determines the scattering partial wave function for a given orbital momentum $\ell$. Here  ${\hat {j_\ell}}$ and ${\hat h}^+_{\ell}$   are the standard Riccati-Bessel  and Riccati-Hankel functions
\cite{Abram}. The exact representation for $v_{\ell}(r,k)$ has a different form depending on  whether the value of $r$ is {\it in} or {\it out} the interval $0<r\le R$ \cite{Ford}. For $r\in (0,R]$
one obtains 
\begin{equation}
v_{\ell}(r,k)=a_{R \ell}\,F_{\ell}(\eta,kr).
\label{vinR}
\end{equation}
For $r\ge R$ the solution $v_{\ell}$ takes the form
\begin{equation}
v_{\ell}(r,k)={{\hat {j_\ell}}}(kr) + A_{R  \ell}\, {\hat h}^+_{\ell}(kr).
\label{voutR}
\end{equation}
At $r=R$ both the function $v_{\ell}$ as well as its first derivative have to be continuous
 in $r$, i.e.,
$\partial_r^n v_{\ell}(R-0,k)=\partial_r^n v_{\ell}(R+0,k)$, $n=0,1$. These conditions yield
\begin{equation}
a_{R \ell}=W_R\{{\hat {j_\ell}}\ ,{\hat h}^+_{\ell}\}/W_R\{F_{\ell}\,,{\hat h}^+_{\ell}\}
,
\label{a_R}
\end{equation}
and
\begin{equation}
A_{R \ell}=W_R\{{\hat {j_\ell}}\,,F_{\ell}\}/W_R\{F_{\ell}\,,{\hat h}^+_{\ell}\},
\label{A_R}
\end{equation}
where $W_R\{f,g\}$ is the Wronskian $f(r)g'(r)-f'(r)g(r)$ that is calculated at $r=R$.
The phase shift $\delta_{R \ell}$ is then determined 
by the standard representation
of the scattering amplitude $A_{R  \ell}$
\begin{equation}
A_{R  \ell}=\frac{\rme^{\rmi 2 \delta_{R  \ell}}-1}{2\rmi}~. 
\label{delta_R}
\end{equation}
It is seen from (\ref{a_R}) and (\ref{A_R}) that the phase shift $\delta_{R\ell}$ can also be given by the argument of
the amplitude $a_{R\ell}$
\begin{equation}
\delta_{R\ell}=  \arg a_{R\ell}.
\label{delta_Rl}
\end{equation}
Using the asymptotics of Riccati-Bessel functions as $kR\gg\ell(\ell+1)$ and
the asymptotics of the regular Coulomb function as $kR\gg \ell(\ell+1)+\eta^2$, one obtains the  asymptotics of
$a_{R\ell}$ 
\begin{equation}
a_{R \ell}\sim \rme^{\rmi(\sigma_{\ell}-\eta\log2kR)}.
\label{a_Ras}
\end{equation}
Therefore, the asymptotics of the phase shift $\delta_{R \ell}$ when $kR\gg \ell(\ell+1)+\eta^2$ reads \cite{Ford}
\begin{equation}
\delta_{R \ell}\sim \sigma_{\ell}-\eta\log 2kR. 
\label{delta_Ras}
\end{equation}
The above procedure describes how
the partial waves $v_{\ell}(r,k)$ can then be constructed. Then
the wave function $v(r,u,k)$ is  given by the infinite sum over momenta $\ell$
\begin{equation}
v(r,u,k)=\frac{1}{kr}\sum_{\ell=0}^{\infty}(2\ell+1)\rmi^{\ell}v_{\ell}(r,k)P_{\ell}(u).
\label{v}
\end{equation}
This function satisfies the three dimensional Schr\"odinger equation (\ref{SE})
with the potential $V_R$ taken instead of $V_{\rmc}$. 

Before proceed further we would like to point out that  the convergence of the partial wave series for the scattering
solutions should be considered with care, especially for the case of
long range potentials \cite{Taylor}.
The most reliable method is by using distribution theory.
Consider an infinitely differentiable test function $f(u)\in C^{\infty}(-1,1)$. By multiplying
both sides of (\ref{v}) with $f(u)$ and integrating over $u$ we obtain
\begin{equation}
\int_{-1}^{1}du\, v(r,u,k)f(u)=\frac{1}{kr}\sum_{\ell=0}^{\infty}(2\ell+1)\rmi^{\ell}v_{\ell}(r,k)\int_{-1}^{1}du\,P_{\ell}(u)f(u).
\label{v-distr}
\end{equation}
Introducing the Fourier coefficients with respect to the Legendre polynomials
\begin{equation}
f_{\ell}=(2\ell+1)\int_{-1}^{1}du\,P_{\ell}(u)f(u),
\label{Fourier f}
\end{equation}
equation  (\ref{v-distr}) can be rewritten in the following form 
\begin{equation}
\int_{-1}^{1}du\, v(r,u,k)f(u)=\frac{1}{kr}\sum_{\ell=0}^{\infty}\rmi^{\ell}v_{\ell}(r,k)f_{\ell}.
\label{dist-pw}
\end{equation}
The series on the right hand side is absolutely and uniformly convergent in the interval $0< r< \infty$,
since the set of Fourier coefficients $f_{\ell}$   forms itself the absolutely convergent series and
the terms of the series (\ref{dist-pw}) can be estimated as 
\begin{equation}
|v_{\ell}(r,k)f_{\ell}|\le C|f_{\ell}|,
\label{vf estim}
\end{equation}
where $C$ is some constant. Thus, the leading term of the asymptotics of the series
when $ r\to \infty$ is now determined by the asymptotics of a certain number of coefficients $v_{\ell}(r,k)$,
whilst the tail of the series is negligible. The detailed description  of using such an
approach for the partial wave series summation can be found in \cite{Taylor}.
In the
following discussion we treat the partial series in the sense described above
while assuming implicitly formulae such as (\ref{v-distr}, \ref{dist-pw}). We also extend this
technique to the operators acting on the square integrable functions of the angular variable $u$.

\subsection{Properties of the solution for $r\le R$}
For  $r\le R$ the equation (\ref{v}) yields
\begin{equation}
v(r,u,k)= \frac{1}{kr}\sum_{\ell=0}^{\infty}(2\ell+1)\rmi^{\ell}a_{R\ell}F_{\ell}(\eta,kr)P_{\ell}(u).
\label{vrlessR}
\end{equation}
The right hand side of (\ref{vrlessR}) is the series in Legendre polynomials \cite{Abram}.
The polynomials $P_{\ell}$  form an orthogonal and complete set of functions on the interval $(-1,1)$ with
respect to the scalar product 
\begin{equation}
\langle f|g\rangle=\int_{-1}^{1} \rmd u\, {{f^*(u)}}g(u),
\label{sprod}
\end{equation}
where the asterisk indicates the complex conjugate. The orthogonality and completeness conditions  for $P_{\ell}$ are
\begin{equation}
\int_{-1}^{1} \rmd u\, P_{\ell}(u) P_{\lambda}(u)= \frac{2}{2\ell+1}\delta_{\ell \lambda},
\label{ortP}
\end{equation}
\begin{equation}
\sum_{\ell=0}^{\infty}\frac{2\ell+1}{2} P_{\ell}(u)P_{\ell}(u')= \delta(u-u').
\label{compP}
\end{equation}
This set provides a basis for the ${\cal L}= L_2(-1,1)$ space of square integrable functions  on the interval $(-1,1)$,
with (\ref{sprod}) as the inner product and  with $\|f\|=\langle
f|f\rangle^{1/2}$ as the norm.
In the following derivation the elements of $\cal L$ will be denoted as vectors, e.g. $|f\rangle$.
With this notation  (\ref{ortP}) and (\ref{compP}) take the
abbreviated form
\begin{equation}
\langle P_{\ell}|P_{\lambda}\rangle = \frac{2}{2\ell+1}\delta_{\ell \lambda},
\end{equation}
\begin{equation}
\sum_{\ell=0}^{\infty}\frac{2\ell+1}{2} |P_{\ell}\rangle \langle P_{\ell}|= {\bf I}.
\end{equation}
Here ${\bf I}$ denotes the unit operator in $\cal L$.  The equation (\ref{vrlessR}) now reads
\begin{equation}
|v(r,k)\rangle=\frac{1}{kr}\sum_{\ell=0}^{\infty}(2\ell+1)\rmi^{\ell}a_{R\ell}F_{\ell}(\eta,kr)|P_{\ell}\rangle.
\label{vabbr}
\end{equation}
Here $|v(r,k)\rangle\in {\cal L}$ represents $v(r,u,k)$ as a function of $u$. From the form of (\ref{vabbr}),
this can be recast into 
\begin{equation}
\begin{array}{l}
|v(r,k)\rangle=\sum\limits_{\ell=0}^{\infty}a_{R\ell}\, \rme^{-\rmi\sigma_{\ell}}\frac{2\ell+1}{2}|P_{\ell}\rangle\langle P_{\ell}|\,
\times \\
\frac{1}{kr}\sum\limits_{\lambda=0}^{\infty}(2\lambda+1)\rmi^{\lambda}\rme^{\rmi\sigma_{\lambda}}F_{\lambda}(\eta,kr)
   |P_{\lambda}\rangle.
\end{array}
\label{A*Psi}
\end{equation}
By comparing this equation to (\ref{pwcwf}) one identifies the right hand side of (\ref{A*Psi}) with the action of the operator
\begin{equation}
{\bf a}_{R} =\sum_{\ell=0}^{\infty}a_{R\ell}\, \rme^{-\rmi\sigma_{\ell}}\frac{2\ell+1}{2}|P_{\ell}\rangle\langle P_{\ell}|
\label{A_Roper}
\end{equation}
on the Coulomb wave function, which in $\cal L$ is represented by the vector
\begin{equation}
|\psi_{\rmc}(r,k)\rangle=\frac{1}{kr}\sum_{\lambda=0}^{\infty}(2\lambda+1)\rmi^{\lambda}\rme^{\rmi\sigma_{\lambda}}
F_{\lambda}(\eta,kr)   |P_{\lambda}\rangle.
\label{psi_c-vec}
\end{equation}
Thus, we have obtained the central focus of this part of the derivation, which establishes the relation between
the solution of the Schr\"odinger equation with the sharply cut-off potential
$V_R$ and the Coulomb wave function for $r\le R$. It has the form 
\begin{equation}
|v({r,k})\rangle={\bf a}_{R}\, |\psi_{\rmc}(r,k)\rangle .
\label{v=Apsi}
\end{equation}
The inverse identity also holds true, yielding
\begin{equation}
|\psi_{\rmc}(r,k)\rangle={\bf a}^{-1}_{R}\, |v({r,k})\rangle.
\label{psi=Amv}
\end{equation}
The exact form of ${\bf a}^{-1}_{R}$  is easily calculated from the orthogonality and completeness of Legendere polynomials
\begin{equation}
{\bf a}^{-1}_{R} =\sum_{\ell=0}^{\infty}a^{-1}_{R\ell}\, \rme^{\rmi\sigma_{\ell}}\frac{2\ell+1}{2}|P_{\ell}\rangle\langle P_{\ell}|.
\label{A_Rmooper}
\end{equation}
The representation (\ref{a_R}) guarantees $|a_{R,\ell}|\ne 0$ and consequently the inverse is correctly defined by
(\ref{A_Rmooper}). The formulae (\ref{v=Apsi}, \ref{psi=Amv}) are
the main results of this section, and they provide us with the representations of the Coulomb wave function in terms of
the wave function $v(r,u,k)$. These representations are valid for an arbitrary
value of the screening radius $R$.

A subsequent reduction of the complexity of the representations (\ref{v=Apsi}) and (\ref{psi=Amv}) can be observed
if $R\to\infty$ by studying the asymptotics of the operator ${\bf a}_R$. The
operator ${\bf a}_R$ can be simplified as $R\to\infty$ with the help of the  asymptotics (\ref{a_Ras}) of the coefficients $a_{R\ell}$.
The natural topology for calculating the asymptotics of ${\bf a}_R$ is the strong operator topology in ${\cal L}$, i.e.
when the asymptotics  of  vectors ${\bf a}_{R}|f\rangle$ in $\cal L$ is
considered as  $R\to\infty$. Construct a vector $|g_R\rangle ={\bf a}_{R}|f\rangle$ with an arbitrary
$|f\rangle\in {\cal L}$ and then represent this vector as a sum of two terms
$|g_{R}\rangle = |g_{LR}\rangle +|g^L_{R}\rangle $, where 
\begin{equation}
\begin{array}{lll}
|g_{LR}\rangle=
\sum\limits_{\ell=0}^{L} a_{R\ell}\,\rme^{-\rmi\sigma_{\ell}}\frac{2\ell+1}{2}|P_{\ell}\rangle\langle P_{\ell}|f\rangle,\\
|g^L_{R}\rangle=
\sum\limits_{\ell=L+1}^{\infty} a_{R\ell}\,\rme^{-\rmi\sigma_{\ell}}\frac{2\ell+1}{2}|P_{\ell}\rangle\langle P_{\ell}|f\rangle.
\end{array}
\label{g_R}
\end{equation}
The second sum can be made arbitrarily small by the choice of $L$. Indeed, due to the known properties of
the Riccati-Hankel function and of the regular Coulomb function \cite{Abram} it is
seen that the coefficients (\ref{a_R}) are bounded $|a_{R\ell}|\le C$, then 
\begin{equation}
\|g^L_{R}\|^2 \le C^2\sum_{\ell=L+1}^{\infty} \frac{(2\ell+1)}{2}|\langle P_{\ell}|f\rangle|^2.
\label{ineq1}
\end{equation}
The Parseval identity for the series in Legendre polynomials
\begin{equation}
\sum_{\ell=0}^{\infty} \frac{(2\ell+1)}{2}|\langle P_{\ell}|f\rangle|^2=\| f\|^2,
\label{Parseval}
\end{equation}
shows that the series on the right hand side of (\ref{Parseval}) is convergent and, as a consequence, its tail 
\begin{equation}
\sum_{\ell=L+1}^{\infty} \frac{(2\ell+1)}{2}|\langle P_{\ell}|f\rangle|^2
\label{tail}
\end{equation}
can be made arbitrarily small if $L$ is taken to be sufficiently large.
From this we conclude that the infinite sum on the right hand side of the inequality  (\ref{ineq1}) can be
made arbitrarily small if $L$ is large enough. Thus, for any small $\epsilon>0$, there exists an integer $L_0>0$
such that for all $L\ge L_0$ the inequality holds 
\begin{equation}
\|g^L_{R}\|^2 \le C^2\epsilon.
\label{ieq0}
\end{equation}
Consider now the vector $|g_{L_0 R}\rangle$. Since $L_0$ is finite there always exists  a value of $R$ such that the
condition $kR \gg L_0(L_0+1)+\eta^2$ is fulfilled. More precisely, the
asymptotics of the Riccati-Hankel function and of the regular Coulomb function can be used for evaluating Wronskians
in (\ref{a_R}), as indicated in (\ref{a_Ras}), to arrive at the
inequality 
\begin{equation}
|a_{RL_0}-\rme^{\rmi(\sigma_{L_0}-\eta\log2kR)}|\le \epsilon^{1/2}.
\label{a_RL0}
\end{equation}
Similar inequalities where $L_0$ is replaced by  $\ell$ for all $\ell\le L_0$ are obviously also true.
With these inequalities we get 
\begin{equation}
\| |g_{L_0 R}\rangle -\rme^{-\rmi\eta\log2kR}{\bf I}_{L_0}|f\rangle \|^2\le \epsilon \|f\|^2,
\label{ieq1}
\end{equation}
where
\begin{equation}
{\bf I}_{L_0}=\sum_{\ell=0}^{L_0}\frac{2\ell+1}{2} |P_{\ell}\rangle \langle P_{\ell}|.
\label{IL0}
\end{equation}
Combining the inequalities (\ref{ieq0}) and (\ref{ieq1}) together with the definition of $|g_R\rangle$ we obtain the final estimate
\begin{equation}
\|({\bf a}_{R}-\rme^{-\rmi\eta\log2kR}{\bf I}_{L_0})|f\rangle\|^2\le \epsilon (C^2+\|f\|^2).
\label{ineqa_R}
\end{equation}
With this estimate, $L_0$ can be extended up to infinity and the inequality with ${\bf I}$ instead of ${\bf I}_{L_0}$
is also valid. Thus, the final result for the asymptotics of the
operator ${\bf a}_R$ is formulated as follows:\\
{\bf Theorem 1}. {\it For any arbitrarily small $\epsilon>0$ there exists  $R$ such that the representation 
\begin{equation}
{\bf a}_R = \rme^{-\rmi\eta\log2kR}{\bf I} +{\bf O}(\epsilon),
\label{theorem}
\end{equation}
holds true. Here the norm of a residual operator ${\bf O}(\epsilon)$ acting on  any vector in $\cal L$ is of the order  $\epsilon$
as in (\ref{ineqa_R}) }.\\

On applying Theorem 1 to the vector $|f\rangle=|\psi_{\rmc}(r,k)\rangle$
the asymptotic form of the relation (\ref{v=Apsi}) is obtained
\begin{equation}
|v(r,k)\rangle = \rme^{-\rmi\eta\log2kR}|\psi_{\rmc}(r,k)\rangle + {\bf O}(\epsilon)|\psi_{\rmc}(r,k)\rangle.
\label{v=aspsi}
\end{equation}
The alternative is given by
\begin{equation}
|\psi_{\rmc}(r,k)\rangle = \rme^{\rmi\eta\log2kR}\left[|v(r,k)\rangle - {\bf O}(\epsilon)|\psi_{\rmc}(r,k)\rangle\right].
\label{v=aspsi-alt}
\end{equation}
These  two last formulae provide the strict basis for the problem of the asymptotic regularization \cite{Ford, Glockle}, which is needed when constructing the three dimensional  Coulomb wave function
from the solution of the Schr\"odinger equation with the screened Coulomb potential $V_R$.

Although the asymptotic regularizing factor $\rme^{\rmi\eta\log2kR}$ is numerical and does not depend on the angular
variable $u$, the general factor ${\bf a}_R$ is the operator in the angular space
$\cal L$. This operator connects the two solutions $v(r,u,k)$ and $\psi_{\rmc}(r,u,k)$ to the same equations for $r\le R$.
This does not lead to a contradiction since the operator ${\bf
a}_R$, or more precisely its extension on the three dimensional space for which we will keep the same notation  ${\bf a}_R$,
commutes with the Hamiltonian ${\bf H}_{\rmc}=-\Delta_{\mbr}+V_{\rmc}(r)$
\begin{equation}
{\bf { a}}_{R} {\bf H}_{\rmc}={\bf H}_{\rmc}{\bf { a}}_{R}.
\label{aH=Ha}
\end{equation}
Thus, if the function $v(r,u,k)$ obeys the Schr\"odinger equation for $r\le R$ 
\begin{equation}
({\bf H}_{\rmc}-k^2)v(r,u,k)=0,
\label{H_c-kk v}
\end{equation}
then  one obtains
\begin{equation}
\begin{array}{lll}
({\bf H}_{\rmc}-k^2)v(r,u,k)&=&({\bf H}_{\rmc}-k^2){\bf a}_R\psi_{\rmc}(r,u,k)=\\
&=& {\bf a}_R({\bf H}_{\rmc}-k^2)\psi_{\rmc}(r,u,k)=0.
\end{array}
\label{Hv}
\end{equation}

For the sake of  completeness it is worth giving the explicit representation of the operator ${\bf a}_R$
acting on the functions of the angular variable 
\begin{equation}
{\bf a}_{R}f (r,u,k)= \int_{-1}^{1}du'\ a_R(u,u')f(r,u',k).
\label{a_R int}
\end{equation}
Here the kernel $a_R(u,u')$ in accordance to (\ref{A_Roper}) is represented  by
\begin{equation}
{a}_{R}(u,u') =\sum_{\ell=0}^{\infty}a_{R\ell}\, \rme^{-\rmi\sigma_{\ell}}\frac{2\ell+1}{2}P_{\ell}(u)P_{\ell}(u').
\label{A_Rkernel}
\end{equation}
This kernel is identical to the function (5) from \cite{Popov} after respective unification of notations.

\subsection{Properties of the solution for $r\ge R$}
For $r\ge R$ the function $v_{\ell}(r,k)$ has the form (\ref{voutR}). The summation over $\ell$ leads to the three dimensional
solution
\begin{equation}
v(r,u,k)=\rme^{\rmi rk u}+v_{\rmsc}(r,u,k)
\label{v3Dout1}
\end{equation}
where the scattered part has the form
\begin{equation}
v_{\rmsc}(r,u,k)=
\frac{1}{kr}\sum_{\ell=0}^{\infty}(2\ell+1)\rmi^{\ell}A_{R  \ell}\, {\hat h}^+_{\ell}(kr)P_{\ell}(u).
\label{vscatt}
\end{equation}
Using standard arguments one arrives at the asymptotics of  $v_{\rmsc}$ as $kr\to\infty$
\begin{equation}
v_{\rmsc}(r,u,k)\sim A_R(u,k)\rme^{\rmi k r}/r
\label{vas1}
\end{equation}
with the partial wave representation for the scattering amplitude $A_R(u,k)$ given by
\begin{equation}
A_R(u,k)=\frac{1}{k}\sum_{\ell=0}^{\infty}(2\ell+1)A_{R\ell}P_{\ell}(u).
\label{scarampl}
\end{equation}
If $R\to\infty$ the amplitude $A_R(u,k)$ can be represented through the Coulomb scattering amplitude. In order to demonstrate this, one needs to use the representation of the amplitude through the
phase shift (\ref{delta_R}) and its asymptotics (\ref{delta_Ras}). {The following representation is first derived}
\begin{equation}
A_{R}(u,k)\sim \frac{1}{k}\sum_{\ell=0}^{\infty}(2\ell+1)\frac{\rme^{\rmi 2(\sigma_{\ell}-\eta\log 2kR)}-1}{2\rmi}
P_{\ell}(u).
\label{A_Ras1}
\end{equation}
By using the standard partial wave expansion of the Coulomb scattering amplitude $A_{\rmc}$
\begin{equation}
A_{\rmc}(u,k)=\frac{1}{k}\sum_{\ell=0}^{\infty}(2\ell+1)\frac{\rme^{\rmi2\sigma_{\ell}}-1}{2\rmi}P_{\ell}(u)
\label{Coulomb-Amp-partial}
\end{equation}
the  representation (\ref{A_Ras1}) can be transformed to the form
\begin{equation}
\begin{array}{l}
A_{R}(u,k)\sim
\rme^{-\rmi 2\eta\log 2kR} A_{\rmc}(u,k)-\\
 -\frac{2}{k}\,\rme^{-\rmi\eta\log 2kR}\sin( \eta\log 2kR)
\sum\limits_{\ell=0}^{\infty}\frac{2\ell+1}{2}P_{\ell}(u).
\end{array}
\label{A_Ras2}
\end{equation}
The sum in the second term of (\ref{A_Ras2}) can easily be evaluated with the help of the completeness of the Legendre polynomials (\ref{compP}) and by taking into account that $P_{\ell}(1)=1$ 
\begin{equation}
\sum_{\ell=0}^{\infty}\frac{2\ell+1}{2}P_{\ell}(u)=\delta(1-u).
\label{delta(1-u)}
\end{equation}
Here the delta-function is understood as in \cite{Taylor}
\begin{equation}
\int_{-1}^{1}du\, \delta(1-u)f(u)=f(1).
\label{delta 1-u def}
\end{equation}
Introducing  (\ref{delta(1-u)}) into the formula (\ref{A_Ras2}) 
we obtain  the final form of the asymptotics of the amplitude (\ref{scarampl})
\begin{equation}
A_{R}(u,k)\sim \rme^{-2\rmi\eta\log 2kR} A_{\rmc}(u,k)
- \frac{2}{k}\,\rme^{-\rmi\eta\log 2kR}\sin( \eta\log 2kR)\,\delta(u-1).
\label{A_Rassfinal}
\end{equation}
This is one of the main results of this section. One can recognize in the first term of (\ref{A_Rassfinal})
the regularization factor, which was derived in \cite{Taylor}. The second term with
the strong delta function singularity was not known until now and, as demonstrated in the analysis shown above,
its exact form can be obtained  by the accurate summation of the all partial terms.

\section{The scattering problem for the potential $V^R$}
In this section, the approach detailed in paper \cite{VEYY} is followed to construct the solution to the partial wave equation
\begin{equation}
\left( -\frac{\rmd^2}{\rmd r^2}+\frac{\ell(\ell+1)}{r^2}+V^R(r)-k^2 \right)w_{\ell}(r,k)=0
\label{pweV_R1}
\end{equation}
for the potential $V^R$.
The exact representation for $w_{\ell}(r,k)$ is of the form
\begin{equation}
w_{\ell}(r,k)=a^R_{\ell}{{\hat {j_\ell}}}(kr)
\label{win}
\end{equation}
provided $r\le R$. For $r\ge R$ this gives 
\begin{equation}
w_{\ell}(r,k)=
\rme^{\rmi\sigma_{\ell}}F_{\ell}(\eta,kr)+A^R_{\ell} u^{+}(\eta,kr).
\label{wout}
\end{equation}
Here $u^{+}(\eta,kr)=\rme^{-\rmi\sigma_{\ell}}(G_{\ell}+\rmi F_{\ell})$ and $G_{\ell}$ is the irregular Coulomb function \cite{Abram}. Similar to (\ref{a_R}, \ref{A_R}), the parameters $ a^R_{\ell}$
and $A^{R}_{\ell}$ should be found by matching  the representations (\ref{win}) and (\ref{wout}) for the function $w_{\ell}(kr)$ and for its first derivative at the point $r=R$. This leads to the
expressions 
\begin{equation}
a^R_{\ell}=\rme^{\rmi\sigma_{\ell}}W_{R}(F_{\ell},u^{+}_{\ell})/W_{R}({\hat{j_{\ell}}},u^+_{\ell}),
\label{a^R}
\end{equation}
\begin{equation}
A^R_{\ell}=\rme^{\rmi\sigma_{\ell}}W_{R}(F_{\ell},{\hat{j_{\ell}}})/W_{R}({\hat{j_{\ell}}},u^+_{\ell}).
\label{A^R}
\end{equation}
The phase shift $\delta^R_{\ell}$ is introduced by the standard representation for the amplitude $A^R_{\ell}$
\begin{equation}
A^R_{\ell}=\rme^{\rmi2\sigma_{\ell}}\frac{\rme^{\rmi2\delta^R_{\ell}}-1}{2\rmi}.
\label{delta^R}
\end{equation}
For large values of $R$ such that $kR\gg \ell(\ell+1)+\eta^2$ the asymptotics of the regular Coulomb function and the Riccati-Bessel function can be used to get the  following asymptotic
representations for amplitudes (\ref{a^R}) and (\ref{A^R}) 
\begin{equation}
a^R_{\ell}\sim \rme^{\rmi\eta\log 2kR},%
\label{a^Ras}
\end{equation}
\begin{equation}
A^R_{\ell}\sim \rme^{\rmi2\sigma_{\ell}} \frac{\rme^{\rmi2(\eta\log 2kR-\sigma_{\ell})}-1}{2\rmi}.
\label{A^Ras}
\end{equation}

The solution to the three dimensional Scr\"odinger equation is given by the sum over the momenta $\ell$ as
\begin{equation}
w(r,u,k)=\frac{1}{kr}\sum_{\ell=0}^{\infty} (2\ell+1)\rmi^{\ell}w^{R}_{\ell}(r,k)P_{\ell}(u).
\label{wu}
\end{equation}
As in the previous section this function takes the special forms on the intervals $0<r\le R$ and $r\ge R$. The $\cal L$ vectors will be used as above for formulating results. For $0<r\le R$ the
vector $|w(r,k)\rangle$ can be represented in terms of the vector $|\psi_{0}(r,k)\rangle$, which represents the plane wave $\psi_{0}(r,u,k)=\rme^{\rmi rku}$, as follows 
\begin{equation}
|w(r,k)\rangle = {\bf a}^R |\psi_{0}(r,k)\rangle.
\label{w}
\end{equation}
The operator ${\bf a}^R$ is represented by
\begin{equation}
{\bf a}^R=\sum_{\ell=0}^{\infty}a^{R}_{\ell}\, \frac{2\ell+1}{2}|P_{\ell}\rangle\langle P_{\ell}|.
\end{equation}
The asymptotics of this operator as $R\to\infty$ can be evaluated in the same way as in the previous section.
The final result should again be understood in the sense of the strong operator
topology in $\cal L$. It reads 
\begin{equation}
{\bf a}^R\sim \rme^{\rmi\eta\log2kR}\, {\bf I}.
\label{aoper^Ras}
\end{equation}
This formula shows that, asymptotically, as $R\to\infty$
\begin{equation}
{\bf a}^R \simeq {\bf a}^{-1}_{R}.
\label{as aa}
\end{equation}

For $r\ge R$ the function $w(r,u,k)$ is given by the expression
\begin{equation}
w(r,u,k)= \psi_{\rmc}(r,u,k) + w_{\rmsc}(r,u,k),
\label{wuout}
\end{equation}
where
\begin{equation}
w_{\rmsc}(r,u,k)=\sum_{\ell=0}^{\infty}(2\ell+1)\rmi^{\ell}A^{R}_{\ell}\, u^{+}_{\ell}(\eta,kr)P_{\ell}(u).
\label{wscat}
\end{equation}
Using standard arguments we obtain the asymptotics of $w_{\rmsc}(r,u,k)$ as $kr\to\infty$
\begin{equation}
w_{\rmsc}(r,u,k)\sim A^R(u,k)\rme^{\rmi (kr-\eta\log 2kr)}/r.
\label{w_sc asympt}
\end{equation}
Here the partial wave decomposition for the amplitude is expressed as
\begin{equation}
A^R(u,k)=\frac{1}{k}\sum_{\ell=0}^{\infty}(2\ell+1)A^R_{\ell}P_{\ell}(u).
\label{A^Rpartialdec}
\end{equation}
If $R\to\infty$ the asymptotics of the amplitude $A^R(u,k)$ should be combined with the Coulomb amplitude $A_{\rmc}(u,k)$ in order to form the total amplitude of the outgoing spherical wave for the
function (\ref{wuout}). In this case we obtain 
\begin{equation}
A_{\rmc}(u,k)+A^R(u,k)\sim \frac{2}{k}\, \rme^{\rmi\eta\log2kR}\, \sin(\eta\log 2kR)\, \delta(1-u).
\label{A_c+A^Ras}
\end{equation}
Comparing to the representation (\ref{A_Rassfinal}) for the amplitude $A_R(u,k)$ one finds the relation
\begin{equation}
A^R \simeq - \rme^{2\rmi \eta \log 2kR} A_R,
\label{A^R-A_R}
\end{equation}
which holds for large values of $R$. This relation can be proved independently by comparing the partial wave series for the amplitudes $A^R(u,k)$ and $A_{R}(u,k)$.

The opposite limit as $R\to 0$ has a  certain  interest for the case of the potential $V^{R}$. It is obvious that $V^R\to V_{\rmc}$. The similar effect  can be  expected for the limit of the wave
function $w(r,u,k)\to \psi_{\rmc}(r,u,k)$. The proof is based on the following asymptotics for the coefficients $a^R_{\ell}$ and $A^R_{\ell}$
\begin{equation}
a^{R}_{\ell}\sim  \rme^{\rmi\sigma_{\ell}}C_{\ell}(2\ell+1)!! ,
\end{equation}
\begin{equation}
A^R_{\ell}\sim \rme^{\rmi2\sigma_{\ell}}\frac{C^2_{\ell}\eta}{\ell+1}(kR)^{2\ell+2}.
\end{equation}
Here $C_{\ell}$ is the standard Coulomb normalization factor \cite{Abram}.
It is easy to see that asymptotically as $R\to 0$
\begin{equation}
a^R_{\ell}{\hat {j_{\ell}}}(kr)\sim \rme^{\rmi\sigma_{\ell}}F_{\ell}(\eta,kr)
\end{equation}
for all $0<r\le R$. When $R\to 0$ the function $u^{+}(\eta,kr)$ becomes singular as $r\to R$
\begin{equation}
u^{+}(\eta,kr)\propto r^{-l}. 
\end{equation}
At the same time the amplitude $A^R_{\ell}$ behaves as
\begin{equation}
A^R_{\ell}\propto 
R^{2(l+1)}. 
\end{equation}
Hence, for all $R\le r< \infty$ one gets
\begin{equation}
\max_{r\in [R,\infty)} |A^R_{\ell}u^{+}_{\ell}(\eta,kr)| = O(R^{\ell+2}),
\end{equation}
which  shows that the term $A^R_{\ell}u^{+}_{\ell}(\eta,kr)$ vanishes faster than the leading term
$\rme^{\rmi\sigma_{\ell}}F_{\ell}(\eta,kr)= O(r^{\ell+1})$ 
when $r\to R$ and $R\to 0$.
Using these estimates in (\ref{wout}) one readily arrives at the statement
\begin{equation}
w(r,u,k)\sim \psi_{\rmc}(r,u,k)
\end{equation}
when $R\to 0$.

\section{The driven Scr\"odinger equation for the Coulomb scattering problem}
In our recent study \cite{VEYY} we demonstrated that the inhomogeneous partial wave Schr\"odinger equation for the scattered
part of the wave function with purely outgoing boundary
conditions can be constructed and then successfully employed  for solving the scattering problem for long range potentials.
The key element of this approach is  the solution of the partial wave
Schr\"odinger equation for the potential $V^R$. The three dimensional approach can now be formulated with the help of the
solution from the preceding  section. The Hamiltonian ${\bf
H}_{\rmc}$ is represented by
\begin{equation}
{\bf H}_{\rmc}=-\Delta_{\mbr}+V_{R}(r)+V^{R}(r)
\label{Ham_C}
\end{equation}
and the wave function $\psi_{\rmc}$ as
\begin{equation}
\psi_{\rmc}(r,u,k)=w(r,u,k)+\psi_{R}(r,u,k).
\label{psi_c}
\end{equation}
Here $w(r,u,k)$ is the wave function for the $V^R$ potential
\begin{equation}
(-\Delta_{\mbr}+V^R(r)-k^2)w(r,u,k)=0,
\label{Schr-eq-_R}
\end{equation}
constructed in the preceding section. The function $\psi_{R}(r,u,k)$ obeys the inhomogeneous (driven) equation 
\begin{equation}
(-\Delta_{\mbr}+V_{\rmc}(r)-k^2)\psi_{R}(r,u,k)=-V_R(r) w(r,u,k)
\label{sepsi_R}
\end{equation}
and the purely outgoing boundary conditions as $r\to\infty,\ \ r\ge R$
\begin{equation}
\psi_{R}(r,u,k)\sim  {\cal A}_{R}(u,k)\rme^{\rmi(kr-\eta\log 2kr)}/r.
\label{psi_R}
\end{equation}
It is seen from the definition that the amplitude ${\cal A}_R$ is given in terms of  $A^R$ through (\ref{A^Rpartialdec}) by 
\begin{equation}
{\cal A}_{R}=-A^R.
\label{Cal A_R}
\end{equation}
For $r\le R$ it is useful to employ the interpretation of the functions as vectors in $\cal L$. In this notations $\psi_{R}$
takes the form 
\begin{equation}
|\psi_{R}(r,k)\rangle =|\psi_{\rmc}(r,k)\rangle -{\bf a}^R|\psi_{0}(r,k)\rangle .
\label{ps}
\end{equation}
and the equation (\ref{sepsi_R}) becomes
\begin{equation}
({\bf H}_{\rmc}-k^2)|\psi_{R}(r,k)\rangle=-{\bf a}^R V_R|\psi_{0}(r,k)\rangle.
\label{sepsi_RL}
\end{equation}
Multiplying (\ref{ps}) by $({\bf a}^R)^{-1}$  and using (\ref{psi=Amv}),  one arrives at the representation
$$
({\bf a}^R)^{-1}|\psi_{R}(r,k)\rangle = ({\bf a}_R{\bf a}^R)^{-1}|v(r,k)\rangle-|\psi_{0}(r,k)\rangle,
$$
which can be reduced using the asymptotic relation (\ref{as aa}) for large values of $R$ to
\begin{equation}
({\bf a}^R)^{-1}|\psi_{R}(r,k)\rangle\simeq  |v(r,k)\rangle-|\psi_{0}(r,k)\rangle.   
\label{a^R-1 asymp}
\end{equation}
By its construction, the function
\begin{equation}
|\phi_{R}(r,k)\rangle =({\bf a}^R)^{-1}|\psi_{R}(r,k)\rangle,
\label{phi_R}
\end{equation}
obeys the equation
\begin{equation}
({\bf H}_{\rmc}-k^2)|\phi_{R}(r,k)\rangle=-V_R|\psi_{0}(r,k)\rangle.
\label{phi_R-eq}
\end{equation}
Following (\ref{a^R-1 asymp}), the function $\phi_{R}(r,k)$ for $r\le R$ and large values of $R$  can be represented as
\begin{equation}
|\phi_{R}(r,k)\rangle\simeq |v(r,k)\rangle-|\psi_{0}(r,k)\rangle.
\label{phi_R sim}
\end{equation}
Equation (\ref{phi_R-eq}) is the desired three-dimensional driven equation, which can be used for solving the Coulomb
scattering problem by the complex rotation method. The formulation of the
scattering problem on the basis of the equation (\ref{phi_R-eq}) obeys two necessary conditions, which are needed for the
application of the complex rotation method, i.e. i) the solution
$\phi_{R}(r,u,k)$ has the purely outgoing asymptotics 
\begin{equation}
\phi_{R}(r,u,k)\sim ({\bf a}^R)^{-1}{\cal A}_{R}\  \rme^{\rmi(kr-\eta\log 2kr)}/r
\label{phi_R asymp}
\end{equation}
and ii) the inhomogeneous term in the right hand side of (\ref{phi_R-eq}) vanishes outside of the sphere of the radius $R$.

\section{Conclusion}
New results on the structure of the solutions to the three dimensional Schr\"odinger equation for the sharply cut-off
Coulomb potential have been derived. For the potential $V_R$,  which
coincides with the Coulomb potential for all $r\le R$,  it was found that the wave function is proportional to the Coulomb
wave function up to an operator factor. This operator acts as an integral
operator over the spherical angular variable. The operator ${\bf a}_R$ is reduced to the multiplication by the constant
$\rme^{-\rmi2\eta\log2kR}$ only asymptotically as $R\to \infty$. This result
clarifies the domain of validity for unjustified assumptions about the proportionality factor, which was taken
as a constant in \cite{Ford} and \cite{Glockle}. The asymptotic
representation of the scattering amplitude for the $V_R$ potential in the case where $R\to\infty$ in addition to the standard
term $\rme^{-\rmi2\eta\log2kR}A_{\rmc}$  also contains
the extra term (\ref{A_Rassfinal}).
It has  fast oscillations as $R\to\infty$ and a strong  delta-functional singularity in the forward scattering direction.
To the best of our knowledge this formula
has been derived here for the first time. The representation for the scattering amplitude obtained in the recent
paper \cite{Glockle} was derived from the incorrect form of the wave function
in the region $r\le R$ \cite{Popov} and cannot be considered as a contra result.
The complete solution for the scattering problem for the potential $V^R$ is given in this paper for the first time.
The formula (\ref{A^R-A_R}) supports the complementary character of the two potentials in the sense
that $V^R=V_{\rmc}-V_R$ and should be considered as the fact of the self consistency of our
treatment. The three dimensional formulation of the driven Schr\"odinger equation,
given in section four, opens the way for forthcoming applications in the three body systems along the line
given in \cite{ElanderFBS}.

\ack

SLY and EY are grateful to Stockholm University for travel and support made possible under the bilateral
agreement between Stockholm University and St Petersburg State
University. The work of SLY and EY was partly supported by the Russian Foundation for Basic Research under the grant 08-02-01115-a.
This work was supported in part by grants from the Swedish National
Research Council.

\end{document}